\documentclass[aps,prl,twocolumn,amsmath,amssymb]{revtex4}
\usepackage{graphicx} 
\usepackage{color}
\definecolor{Red}{rgb}{1.0,0.0,0.0}

\begin{document}

\title{Viscoelastic multiscaling in immersed networks}

\author{J. L. B. de Ara\'ujo}
\affiliation{Departamento de F\'{i}sica, Universidade Federal do 
Cear\'{a}, 60451-970 Fortaleza, Cear\'{a}, Brazil}

\author{J. S. de Sousa}
\affiliation{Departamento de F\'{i}sica, Universidade Federal do 
Cear\'{a}, 60451-970 Fortaleza, Cear\'{a}, Brazil}

\author{W. P. Ferreira}
\affiliation{Departamento de F\'{i}sica, Universidade Federal do 
Cear\'{a}, 60451-970 Fortaleza, Cear\'{a}, Brazil}
  
\author{C. L. N. Oliveira}
\email{lucas@fisica.ufc.br}
\affiliation{Departamento de F\'{i}sica, Universidade Federal do 
Cear\'{a}, 60451-970 Fortaleza, Cear\'{a}, Brazil}

\begin{abstract}
Rheological responses are the most relevant features to describe soft
matter. So far, such constitutive relations are still not well
understood in terms of small scale properties, although this knowledge
would help the design of synthetic and bio-materials.  Here, we
investigate, computational and analytically, how mesoscopic-scale
interactions influence the macroscopic behavior of viscoelastic
materials.  We design a coarse-grained approach where the local
elastic and viscous contributions can be controlled.  Applying
molecular dynamics simulations, we mimic real indentation assays.
When elastic forces are dominant, our model reproduces the hertzian
behavior.  However, when friction increases, it restores the Standard
Linear Solid model.  We show how the response parameters depend on the
microscopic elastic and viscous contributions. Moreover, our findings
also suggest that the relaxation times, obtained in relaxation and
oscillatory experiments, obey a universal behavior in viscoelastic
materials.
\end{abstract}

\maketitle

One big challenge in science and engineering is the so-called
multiscale modeling, namely, how constitutive relations of a material
depend on its smaller scale interaction and
composition~\cite{Yip2013,White2009}.  Such emergence problem, as
described by P. W. Anderson~\cite{Anderson1972}, is even more
remarkable in soft matter, where mesoscopic structures combine the
atomistic and macroscopic frameworks and are responsible to the
material elasticity and viscosity~\cite{Doi2013,Praprotnik2008}. In
fact, the link among features of different scales forming the matter
is highly nontrivial and subjected to intense research~\cite{Qu2011}.
The seminal work of H. Hertz about mechanical contacts, for instance,
is still used nowadays to measure elastic properties of materials by
analyzing how samples are deformed under applied
stresses~\cite{Hertz1881}. Although it is one of the most used models
to investigate the stiffness of a material, it fails to relate the
macroscopic behavior with its inner parts. Besides, well-known
analytical approaches such as the Maxwell, Kelvin-Voigt, and Standard
Linear Solid models apply circuit analogies to propose a simple manner
of how elastic and viscous terms mix up in viscoelastic
materials~\cite{Lakes2017}.  Again, these rheological models neglect
any downscaling analysis.

Currently, many sophisticated methods have been developed to study the
viscoelasticity of materials. Nanoindentation experiments, for
instance, such as those with Atomic Force Microscopy, are extensively
applied to investigate the mechanical properties at micrometer
scale. Besides condensed matter, such technique has also been used to
investigate soft materials~\cite{Lin2008,Radmacher1997,Rebelo2013}.
The way the sample responses to external stresses depends on its
elastic and viscous terms.  However, the complex structure and
composition in soft matter systems make it challenging to investigate
accurately the distribution of applied stresses through their
interior. Many soft matter systems hold large colloidal aggregations
or long (linear or cross-linked) polymeric chains, which have an
essential mechanical role~\cite{Doi2013}. In living cells, for
instance, the cytoskeleton and the cytoplasmic fluid are constantly
exchanging momentum with the extracellular
surroundings~\cite{Gupta2015,Moeendarbary2013}.  The knowledge of why
these organelles influence the cell stiffness differently in healthy
and sick cells, may lead to treatments for several
diseases~\cite{Sousa2020}.
 
To investigate the link between mesoscopic and macroscopic features,
we employ a molecular coarse-grained approach to reproduce rheological
behavior of soft matter. We design a viscoelastic material composed of
a particle-spring network immersed in a viscous medium so that the
contribution of elastic and viscous interactions can be controlled at
the mesoscopic level. This immersed network model is compatible with
suspended polymer chains, colloidal aggregations, and other
load-bearing structures as commonly found in soft
materials~\cite{Doi2013}. Although real materials unlikely present
purely quadratic potentials, for small deviations, molecular
interactions displaying a potential well can be described
appropriately by a spring-like interaction. For instance, the
equivalent spring constant of the Lennard-Jones interatomic potential
can be calculated as $72\epsilon_0/\ell_0^2$, where $\epsilon_0$ is
the height of the potential well and $\ell_0$ the equilibrium distance
between molecules~\cite{Kleppner2014}. Moreover, the computational
model we develop here is easily changed to support a more realistic
potential in order to study more complex materials. We also consider
the spring network as a regular lattice. Although complex structures
may influence rheological properties of a material, simple geometries
have been successfully applied to a large number of problems in
mesoscopic models of condensed~\cite{Oliveira2012,Moreira2012} and
soft materials~\cite{Oliveira2014,Alves2016,Dias2012}.

\begin{figure}[t]
\begin{center}
\includegraphics[width=0.8\columnwidth]{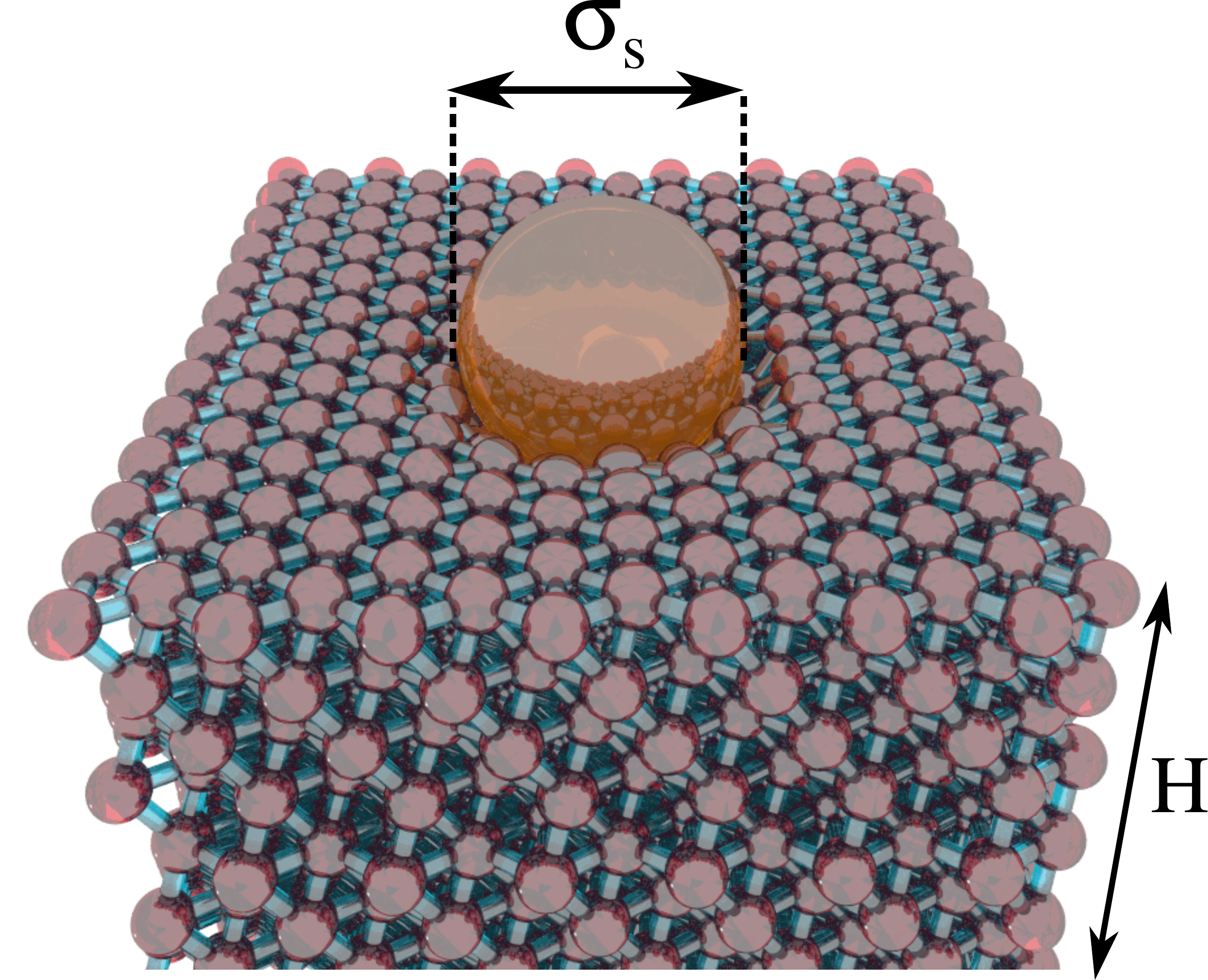}
\end{center}
\caption{Particle-spring network arranged in an FCC (face-centered
  cubic) lattice with height $H$ and immersed in a viscous medium (not
  shown in the figure). Particles are bonded to their 12 nearest
  neighbors.  A spherical indenter of diameter $\sigma_s$ is used to
  press down the top surface of the network.}
\label{fig1}
\end{figure}

Our computational model consists of $N$ spherical particles of
diameter $\sigma$ and mass $m$ arranged in a face-centered cubic (FCC)
lattice with a height given by $H$ and a bottom and top plane given by
$20\sigma\times 20\sigma sin(\pi/3)$.  Every particle interacts with
its 12 nearest neighbors by a spring potential. This elastic network
is immersed in a medium of viscosity $\mu$. Viscous effects are taken
into account by considering friction between the particles and the
medium. The network is indented by a hard sphere of diameter
$\sigma_{s}$, as shown in Fig.~\ref{fig1}. This spherical indenter
moves down with constant speed until a maximum indentation depth,
$\delta_0$, is achieved. To avoid non-linearities, we consider
$\delta_0=\sigma$, i.e., the maximum indentation is equal to the
thickness of one layer of particles. Once in contact, the indenter
applies a stress onto the particles. The bottom layer of the network
is in contact to a hard substrate, where particles cannot move down
but are free to slide horizontally.

The equation of motion of the $i_{th}$ particle is given by the
following Langevin-like equation~\cite{Langevin1908,Langevin}
\begin{equation}
m\frac{d^2 \vec{r}_i}{d t^2} = - \gamma \vec{v}_i - \nabla U_i,
\label{eq:eq1}
\end{equation}
where $\vec{r}_i$ and $\vec{v}_i$ are the position and velocity
vectors of particle $i$, respectively. The term $\gamma \vec{v}_i$,
known as Stoke's law, describes the drag force acting on spherical
bodies, where the coefficient of friction is given by
$\gamma=3\pi\sigma\mu$~\cite{Batchelor1967}.  The potential $U_i$ of
particle $i$ due to the neighboring particles and the indenter is
given by
\begin{equation}
U_i=\sum_{j}\frac{k_{ij}(r_{ij}-\ell)^{2}}{2} 
+ \epsilon \left(\frac{\sigma}{r_{is}-(\sigma_s-\sigma)/2}\right)^{\alpha},
\label{eq:Ui}
\end{equation}
where $r_{ij}=|\vec{r}_i-\vec{r}_j|$ is the distance and $\ell$ the
equilibrium distance between the centers of neighboring particles $i$
and $j$. $k_{ij}$ is the spring constant of the bond between these two
particles.  The last term of Eq.~\ref{eq:Ui} applies only to surface
particles, when they are in contact with the indenter.
$r_{is}=|\vec{r}_i-\vec{r}_s|$ is the distance between the center of
particle $i$ and the center of the indenter, $\epsilon$ is a constant
of energy, and $\alpha$ regulates the hardness of the indenter.  The
particles and the indenter exclusively interact through this
hard-sphere potential with a high value of $\alpha$. See in
Ref.~\cite{param} the constants used in this work.

We solve the equations of motion in (\ref{eq:eq1}) through molecular
dynamics simulations with a time integration done with the Velocity
Verlet algorithm, and periodic boundary conditions applied to the
horizontal plane~\cite{Rapaport2004,Araujo2017}. The contact force $F$
is the sum of all collisions on the indenter, computed at each time as
the indenter slowly presses down the sample. $F$ increases with the
indentation depth, $\delta$, since more collisions occur on the
indenter.  These collisions cause a fluctuation in $F$, but an
equilibrium state is reached in around $10^5$ time steps. After
equilibrium, we perform an additional $2\times 10^5$ time steps to
average the quantities of interest.

\begin{figure}[t]
\begin{center}
\includegraphics[width=0.85\columnwidth]{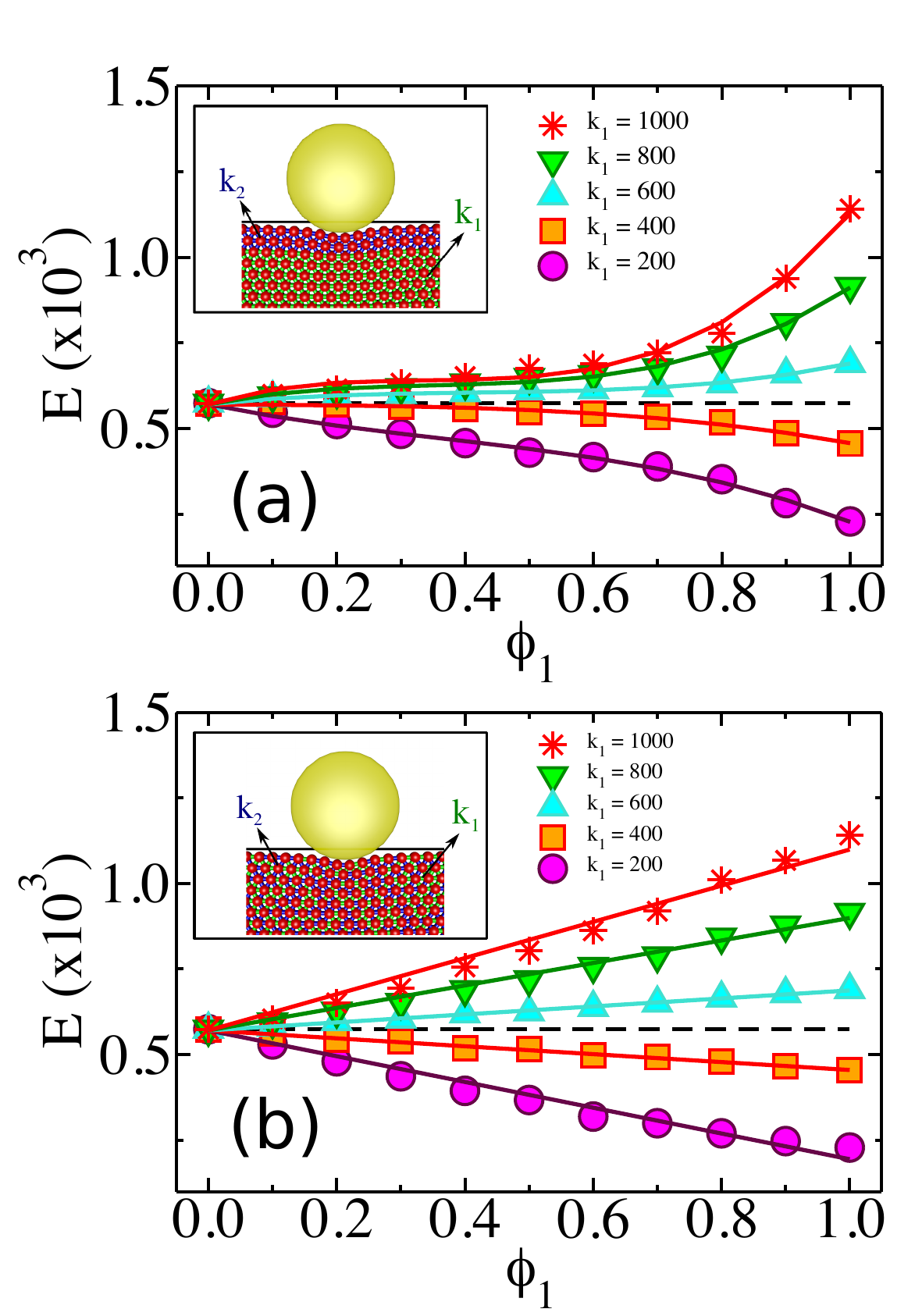}
\end{center}
\caption{The effective modulus of elasticity, $E$, as a function of
  the fraction of $k_1$-bonds in heterogeneous elastic networks,
  $\phi_1$, for $k_2=500$ and several values of $k_1$.  The
  double-layered network is in shown in (a) while the binary random
  network in (b).  The {\it inset} of each graph shows how $k_1$ and
  $k_2$ are distributed in the network.  The black dashed lines in
  each graph represent the elastic modulus of homogeneous networks,
  for $k_1=k_2=500$, with $E=570.75$. Cubic functions and straight
  lines, shown in colored solid lines, are used to fit each set of
  points in (a) and (b), respectively.}
\label{fig2}
\end{figure}

Before viscoelastic materials, we explore the elasticity in networks
without local friction ($\gamma=0$). Initially, we study homogenous
networks, where all bonds have the same spring constant, $k_{ij}=k$,
and later we investigate the role of heterogeneities of local
elasticity in the global stiffness. For simplicity, we consider
heterogeneous networks made of only two types of bonds with spring
constants given by $k_1$ and $k_2$, respectively. Moreover, this
binary network is built in two fashions depending on the distribution
of $k_1$ and $k_2$, namely, the {\it double-layered} and the {\it
  binary random network}.  In the first, two homogenous networks, with
spring constants given by $k_1$ and $k_2$, respectively, are deposited
one on the top of the other (see the {\it inset} in
Fig.~\ref{fig2}(a)).  In the second, $k_{ij}$ is randomly assigned to
$k_1$ or $k_2$ according to probabilities $\phi_1$ and $\phi_2$,
respectively (see the {\it inset} in Fig.~\ref{fig2}(b)). In both
cases, $\phi_1$ and $\phi_2$ also stand to the fraction of each bond
and $\phi_1+\phi_2=1$.  For the limit cases of $\phi_1=0$ and
$\phi_1=1$, both the double-layered and the binary random networks
become homogeneous with spring constant $k_2$ and $k_1$, respectively.

The distribution of stress into an elastic sample, due to an
indentation $\delta$, can be tricky~\cite{Sneddon1965}, however the
force experienced by a spherical indenter is well described by the
Hertz model
\begin{equation}
 F_H = \frac{8\sqrt{2}}{9}E\sqrt{\sigma_s}\delta^{3/2},
\label{eq:hertz}
\end{equation}
where $E$ is the effective elastic modulus of the
material~\cite{Dimitriadis2002} (see Fig.~S1 in the Supplementary
Material).  The contact forces computed in elastic networks perfectly
agree with the Hertz model, where $E$ can be obtained by fitting the
data with Eq.~\ref{eq:hertz}. The results for homogeneous elastic
networks are shown in Fig.~S2 in the Supplementary Material for
different values of $k$.  As expected for this simple network, the
effective modulus of elasticity scales linearly with $k$. In addition,
$E$ is proportional to the sample height in the form $E\propto
H^{-\zeta}$, where $\zeta$ depends on the indenter geometry. Spherical
and flat indenters, for instance, give $\zeta =1/3$ and 1,
respectively. In fact, in the latter case, $E$ represents the Young's
modulus of the material and can be found analytically with springs in
series, while in the other case, $E$ seems to take into account,
besides Young's modulus, the shear modulus of the material.

The effective elastic modulus as a function of $\phi_1$ is shown in
Fig.~\ref{fig2}(a), for double-layered networks, and in
Fig.~\ref{fig2}(b), for binary random networks, for $k_2=500$ and
different values of $k_1$.  $E$ decreases with $\phi_1$ for $k_1<k_2$,
but increases with $\phi_1$ for $k_1>k_2$, regardless the distribution
of $k_1$ and $k_2$.  In double-layered networks, $E$ is fitted by the
cubic function $E=570.75+a\phi+b\phi^2+c\phi^3$, shown in colored
solid lines. On the other hand, when $k_1$ and $k_2$ are randomly
distributed over the network, $E$ changes according to the straight
line $E=570.75+d\phi$, shown in colored solid lines.  The constants
$a$, $b$, $c$, and $d$ depend on $k_1$. Interestingly, the random
distribution of local stiffness vanishes those polynomial higher order
found in (a).

\begin{figure}[t]
\begin{center}
\includegraphics[width=0.9\columnwidth]{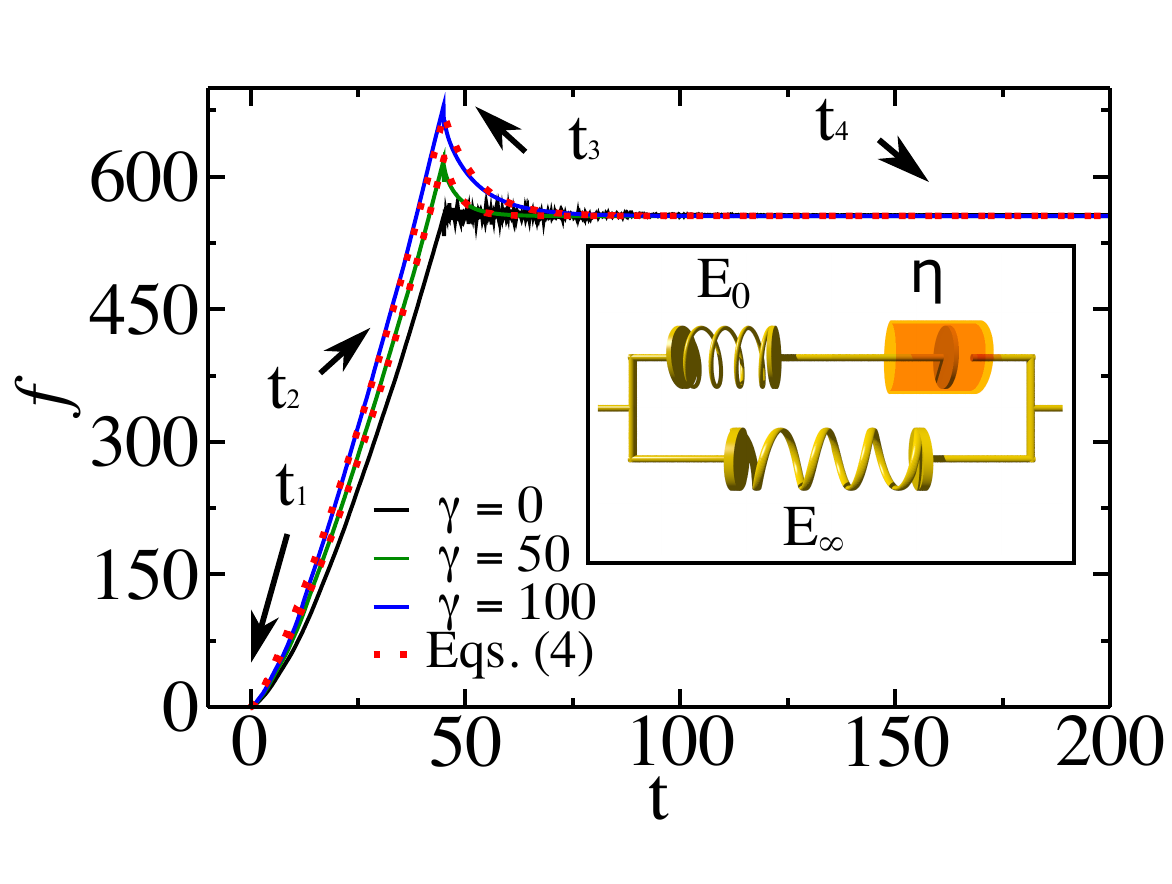}
\end{center}
\caption{Relaxation experiments in viscoelastic networks, for $k=500$
  and different values of $\gamma$. The contact force increases in the
  loading stage and decreases in the dwell stage. The force is
  normalized by geometric constants (see Eq.~4 in the Supplementary
  Material). Our numerical results are compatible with the Standard
  Linear Solid (SLS) model given in Eqs.~\ref{flfd}. A schematic
  illustration of the SLS model is shown in the {\it inset}, with a
  dashpot of viscosity $\eta$ and two springs of elastic moduli $E_0$
  and $E_{\infty}$.}
\label{fig3}
\end{figure}

In order to study the contribution of viscous effects ($\gamma>0$) in
the macroscopic rheological behavior, we perform relaxation and
oscillatory assays in homogenous viscoelastic networks.  In relaxation
experiments, the indenter is initially located above the network and
moving down into the top surface. Before touching the sample, the
force is zero. When the indenter touches the surface particles, at
time $t_1=0$, the force rises until the maximum indentation depth,
$\delta_0$, is achieved, at $t_3$. Figure~\ref{fig3} shows the time
evolution of the normalized contact force, $f$, for $k=500$ and
several values of $\gamma$. The loading time, $\tau_l=t_3-t_1$, is the
time the indenter pushes the sample.  After reaching $\delta_0$, the
indenter stops moving, but the sample particles may continue to move
subjected to friction (the dwell stage).  For elastic $\gamma=0$, the
particles immediately rearrange themself to an equilibrium state and
thus the force becomes nearly constant, except for some noise due to
the perpetual undamped movement of the particles. On the other hand,
in viscoelastic networks, this noise quickly vanishes while the force
decreases continuously to the value of the undamped case, at $t_4$,
since particles need time to relax into the equilibrium state. For
long times, networks with the same elastic properties experience the
same contact force, regardless of the viscous contribution.

Our numerical results are compatible with the Standard Linear Solid
(SLS) model composed of a Maxwell material (i.e., a spring of elastic
modulus $E_0$ in series with a dashpot of viscosity $\eta$) in
parallel to another spring of elastic modulus $E_{\infty}$, see
Fig.~\ref{fig3}. Such analytical model successfully describes
rheological behaviors of a large number of soft materials, such as
polymers~\cite{Plaseied2008}, soft gels~\cite{Okamoto2011}, and living
cells~\cite{Peeters2005,Koay2003}.  The relaxation function of this
model is know as $R(t)=E_{\infty}+E_0e^{-t/\tau}$, where the
relaxation time is given by $\tau=\eta/E_0$.  Initially, the effective
modulus of elasticity is given by the sum of $E_{\infty}$ and $E_0$,
but, for long times, the influence of $E_0$ vanishes, leaving
$E_{\infty}$ to dominate the elasticity of the material, regardless of
the relaxation time. Notice that, $E_{\infty}$ corresponds to the
elastic modulus computed in elastic networks. The analytical force
curves in the SLS model with a spherical indenter, $f(t)$, can be
obtained separately in the loading, $f_l(t)$, and dwell stages,
$f_d(t)$, as follows
\begin{eqnarray}\nonumber
 f_l(t) &=& a_1t^{3/2}+a_2
 \sqrt{t}+a_3\,erf\left(\sqrt{\frac{t}{\tau}}\right)e^{-t/\tau},
 \hspace{0.1cm}t\leq\tau_l,\\ f_d(t)
 &=& a_4+a_5
 e^{-(t-\tau_l)/\tau}+a_6e^{-t/\tau},\hspace{0.7cm}t\geq\tau_l,
 \label{flfd}
\end{eqnarray}
where $erf(t)$ is the error function and the constants are given by
\begin{eqnarray}\nonumber
  a_1 &=& \frac{E_{\infty}}{\tau_l^{3/2}},\,\,\, a_2 =
  \frac{3}{2}\frac{\tau}{\tau_l^{3/2}}E_0,\,\,\, a_3 =
  -\frac{3}{4}\sqrt{\pi}E_0\left(\frac{\tau}{\tau_l}\right)^{3/2},\\\nonumber
  a_4 &=& E_{\infty},\,\,\, a_5 =
  \frac{3}{2}\frac{\tau}{\tau_l}E_0,\,\,\, a_6 =
  a_3\,erf\left(\sqrt{\frac{\tau_l}{\tau}}\right).
\end{eqnarray}
See more details in Section I of the Supplementary Material.

\begin{figure}[t]
 \begin{center}
  \includegraphics[width=0.85\columnwidth]{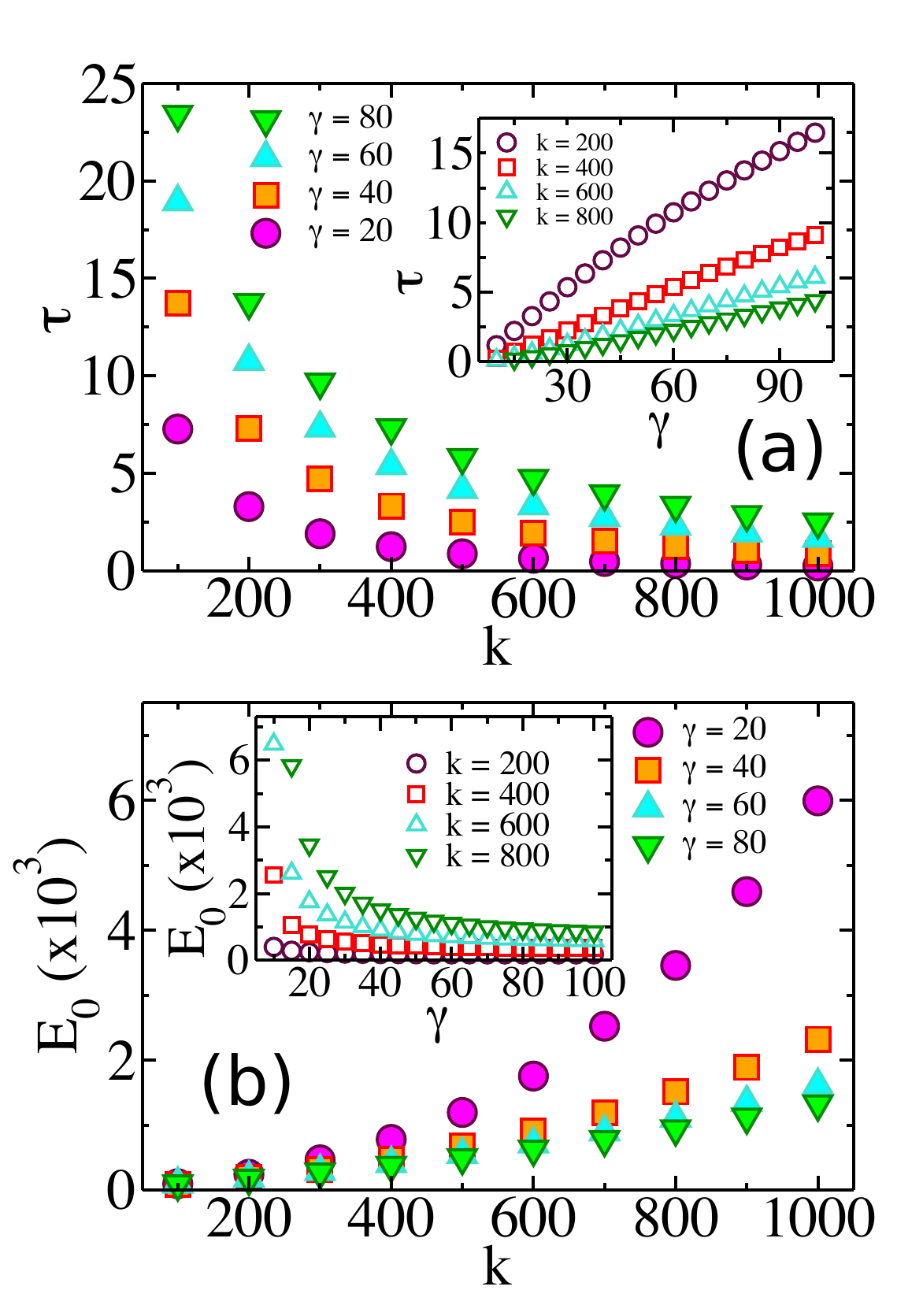}
  \end{center}
  \caption{The macroscopic properties $\tau$ and $E_0$, in (a) and
    (b), respectively, as a function of microscopic parameters, $k$
    and $\gamma$, in viscoelastic networks under relaxation assays.}
  \label{fig4}
\end{figure}

Combining our numerical results with Eqs.~\ref{flfd} allows us to find
the relations between macroscopic ($\tau$ and $E_0$) and microscopic
parameters ($k$ and $\gamma$) in viscoelastic networks.  As shown for
elastic networks, $E_{\infty}$ increases linearly with $k$ and does
not depend on $\gamma$.  However, $\tau$ decreases with $k$ and
increases with $\gamma$ (see Fig.~\ref{fig4}(a)).  On the other hand,
$E_0$ increases with $k$ but decreases with the local dissipation (see
Fig.~\ref{fig4}(b)).

In oscillatory experiments, instead of stoping the indenter after the
maximum indentation is achieved, a sinusoidal strain is imposed with
the indentation depth following the function, $\delta
(t)=\delta_0+\delta_asin(\omega t)$, where $\delta_a$
($\delta_a=\sigma$) is the amplitude and $\omega$ the angular
frequency of oscillation.  The response force $F$ also presents a
periodic behavior but with a time delay that depends on viscoelastic
properties, $F = F_0+F_asin(\omega t + \lambda)$, where $F_0$ is the
force at $\delta_0$, $F_a$ is the amplitude of the force and $\lambda$
is the phase lag between force and indentation. Figure~S3(a) and (b),
in the Supplementary Material, show $\delta(t)$ and $F(t)$,
respectively, for a sample with $k=500$ and three values of $\gamma$.
The phase lag vanishes for $\gamma=0$, as it is expected for elastic
materials, but increases with $\gamma$, for viscoelastic materials.

The tangent of $\lambda$, defined as the ratio between the loss
modulus, $G^{'}=\frac{\delta_a}{F_a}sin(\lambda)$, and the storage
modulus, $G^{''}=\frac{\delta_a}{F_a}cos(\lambda)$, represents a
quantitative way to define the fluidity of viscoelastic materials,
i.e., $\tan(\lambda)<1$ leads to solid-like materials while
$\tan(\lambda)>1$ to fluid-like materials.  High values of $\omega$
favor the fluidity of the material, as shown in panel of
Fig.~\ref{fig5} for several values of $\gamma$. Figure~S3(c) and (d),
in the Supplementary Material, shows color maps of $\tan(\lambda)$ for
more values of $k$ and $\gamma$. We only observe viscoelastic solids
in all cases of $k$ and $\gamma$ considered in this work.  Fluid-like
behavior should be observed for very high values of $\omega$, which
requires a time step too small compared to the time of the experiment.
Moreover, viscoelastic materials may present different relaxation
times regarding of how the sample is deformed. In this work, we
obtained $\tau$ and $\tau_o$ in relaxation and oscillatory conditions,
respectively.  Standard Linear Solids under oscillatory strain leads
to the following behavior~\cite{Lakes2017}
\begin{equation}
\tan(\lambda) = A\frac{\omega\tau_o}{1+\omega^2\tau_o^2},
\label{eq:lambda}
\end{equation}
which is used to fit those numerical data in Fig.~\ref{fig5} in order
to obtain $\tau_o$, where $A$ is a constant of the material.  The main
graph in Fig.~\ref{fig5} shows that $\tau$ and $\tau_o$ depend on each
other by the following exponential relation,
$\tau=0.466(e^{0.08\tau_o}-1)$. Each point in the graph of
Fig.~\ref{fig5}(b) represents a different material, with different $k$
and $\gamma$, but they all collapse on that relation, suggesting some
kind of universal behavior.

In conclusion, we studied macroscopic rheological behaviors in
viscoelastic soft materials in terms of their microscopic elastic and
viscous constituents.  Different indentation experiments were
performed to investigated how the local stiffness and the coefficient
of friction induce different macroscopic behaviors.  Elastic networks
recovers the Hertz model for mechanical contact, where the effective
modulus of elasticity, $E$, is obtained.  In homogeneous elastic
networks, we found that $E\propto H^{-\zeta}$, where $H$ is the
network height and the exponent depends on the indenter geometry. For
spherical and flat indenters, for instance, $\zeta=1/3$ and $\zeta=1$,
respectively. Heterogeneous elastic networks, for simplicity, are
assumed to be binary networks, where each bond holds one of only two
possible values, $k_1$ or $k_2$.  When $k_1$ and $k_2$ are separated
forming layers, $E$ follows a cubic function with $\phi_1$, however,
when $k_1$ and $k_2$ are randomly distributed, the higher order terms
vanish and $E$ changes linearly with $\phi_1$.  Viscoelastic networks
are compatible with the analytical Standard Linear Solid (SLS) model,
which is described by two elastic moduli, $E_0$ and $E_\infty$, and a
relaxation time, $\tau$.  Therefore, our work applies to all those
materials that can be described by the Hertz and SLS models, such as
polymers, gels and living cells.  In oscillatory experiments, we also
computed the phase lag between force and indentation, $\lambda$, and
the relaxation time, $\tau_o$.  We showed how all these macroscopic
properties ($E_0$, $E_\infty$, $\tau$, $\lambda$, and $\tau_o$) depend
on the microscopic parameters.  In addition, the interplay between
different relaxation times, $\tau$ and $\tau_0$, obtained in
relaxation and oscillatory experiments, follows an exponential
behavior, regardless of the mesoscopic interactions. All viscoelastic
networks simulated here collapsed on this curve, suggesting a
universal behavior in viscoelastic materials.

\begin{figure}[t]
\begin{center}
\includegraphics[width=0.85\columnwidth]{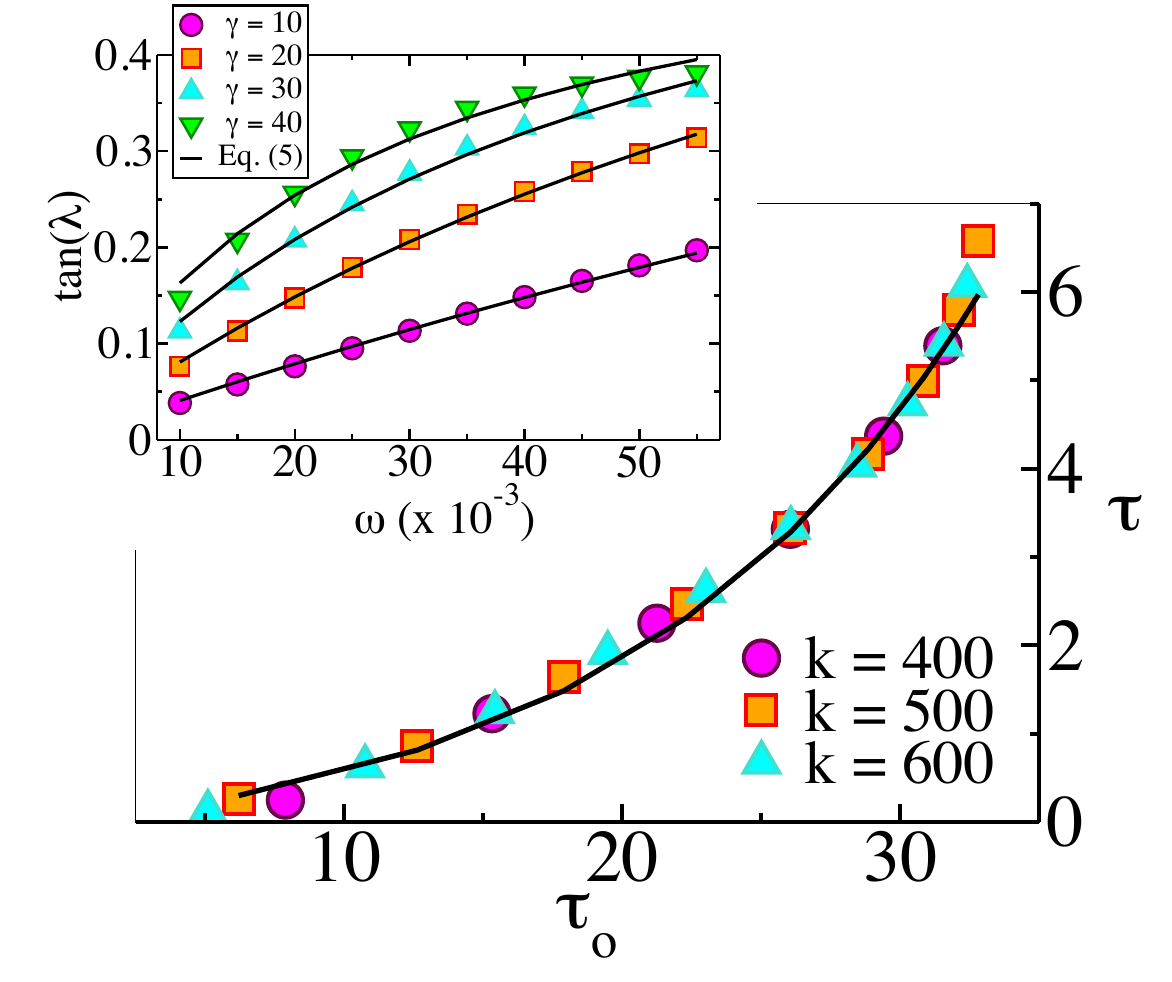}
\end{center}
\caption{In oscillatory experiments, the tangent of the phase lag,
  $\lambda$, as a function of $\omega$, is shown in the panel for
  $k=500$ and different values of $\gamma$. Our results recover in a
  good agreement the behavior of $\tan(\lambda)$ given in
  Eq.~\ref{eq:lambda}, where the relaxation time in oscillatory
  experiments, $\tau_o$, is obtained. The relaxation times, $\tau$ and
  $\tau_o$, obtained in different experiments of deformation, depend
  on each other as follows, $\tau=0.466(e^{0.08\tau_o}-1)$ (shown in
  black solid line), regardless of $k$ and $\gamma$, as shown in the
  main graph.}
\label{fig5}
\end{figure}

\begin{acknowledgments}
  We acknowledge financial support from the Brazilian agencies CNPq,
  CAPES, FUNCAP. This work was also supported by the Serrapilheira
  Institute (grant number Serra-1709-18453).
\end{acknowledgments}


\begin{thebibliography}{100}

\bibitem{Yip2013} S. Yip and M. P. Short, Multiscale materials
  modelling at the mesoscale, Nature Mater. {\bf 12}, 774 (2013).

\bibitem{White2009} R. J. White, G. C. Y. Peng, and S. S. Demir,
  Multiscale modeling of biomedical, biological, and behavioral
  systems (Part 1) [Introduction to the special issue], IEEE Eng. Med
  Biol. Mag. {\bf 28} 12 (2009).

\bibitem{Anderson1972} P. W. Anderson, More is different, Science {\bf
  177}, 393 (1972).

\bibitem{Doi2013} Masao Doi, {\it Soft Matter Physics} (Oxford
  University Press, 2013).

\bibitem{Praprotnik2008} M. Praprotnik, L. D. Site, and K. Kremer,
  Multiscale Simulation of Soft Matter: From Scale Bridging to
  Adaptive Resolution, Annu. Rev. Phys. Chem. {\bf 59}, 545 (2008).

\bibitem{Qu2011} Z. Qu, A. Garfinkel, J. N. Weiss, and M. Nivala,
  Multi-scale modeling in biology: How to bridge the gaps between
  scales?, Prog. Biophys. Mol. Biol. {\bf 107}, 21 (2011).

\bibitem{Hertz1881} H. Hertz, \"Uber die Ber\"uhrung fester
  elastischer K\"orper, J. Reine Angew Mathematik {\bf 95}, 156
  (1881).

 \bibitem{Lakes2017} R. S. Lakes, {\it Viscoelastic Solids} (CRC
   Press, 2017).

\bibitem{Lin2008} D. C. Lin and F. Horkay, Nanomechanics of polymer
  gels and biological tissues: A critical review of analytical
  approaches in the Hertzian regime and beyond, Soft Matter {\bf 4},
  669 (2008).

\bibitem{Radmacher1997} M. Radmacher, Measuring the elastic properties
  of biological samples with the AFM, IEEE Eng. Med. Biol. Mag. {\bf
    16}, 47 (1997).

\bibitem{Rebelo2013} L. M. Rebelo, J. S. de Sousa, J. Mendes Filho,
  and M. Radmacher, Comparison of the viscoelastic properties of cells
  from different kidney cancer phenotypes measured with atomic force
  microscopy, Nanotech. {\bf 24}, 055102 (2013).

\bibitem{Gupta2015} M. Gupta, B. R. Sarangi, J. Deschamps,
  Y. Nematbakhsh, A. Callan-Jones, F. Margadant, R.-M. M\`ege,
  C. T. Lim, R. Voituriez, and B. Ladoux, Nature Commun. {\bf 6}, 7525
  (2015).

\bibitem{Moeendarbary2013} E. Moeendarbary, L. Valon, M. Fritzsche,
  A. R. Harris, D. A. Moulding, A. J. Thrasher, E. Stride,
  L. Mahadevan, and G. T. Charras, Nature Mater. {\bf 12}, 253 (2013).

\bibitem{Sousa2020} J. S. de Sousa, R. S. Freire, F. D. Sousa,
  M. Radmacher, A. F. B. Silva, M. V. Ramos,
  A. C. O. Monteiro-Moreira, F. P. Mesquita, M. E. A. Moraes,
  R. C. Montenegro, and C. L. N. Oliveira, Double power-law
  viscoelastic relaxation of living cells encodes motility trends,
  Sci. Rep. {\bf 10}, 4749 (2020).

 \bibitem{Kleppner2014} D. Kleppner and R. Kolenkow, {\it An
   Introduction to Mechanics} - second edition (Cambridge University
   Press, 2014).

\bibitem{Oliveira2012} C. L. N. Oliveira, A. P. Vieira,
  H. J. Herrmann, and J. S. Andrade, Jr., Subcritical fatigue in fuse
  networks, Europhys. Lett. {\bf 100}, 36006 (2012).
  
\bibitem{Moreira2012} A. A. Moreira, C. L. N. Oliveira, A. Hansen,
  N. A. M. Ara\'ujo, H. J. Herrmann, and J. S. Andrade, Jr.,
  Fracturing Highly Disordered Materials, Phy. Rev. Lett. {\bf 109},
  255701 (2012).

\bibitem{Oliveira2014} C. L. N. Oliveira, J. H. T. Bates, and B. Suki,
  A network model of correlated growth of tissue stiffening in
  pulmonary fibrosis, New J. Phys. {\bf 16}, 065022 (2014).

\bibitem{Alves2016} C. Alves, A. D. Ara\'ujo, C. L. N. Oliveira,
  J. Imsirovic, E. Bartol\'ak-Suki, J. S. Andrade, and B. Suki,
  Homeostatic maintenance via degradation and repair of elastic fibers
  under tension, Sci. Rep. {\bf 6}, 27474 (2016).

\bibitem{Dias2012} C. S. Dias, N. A. M. Ara\'ujo, and M. M. Telo da
  Gama, Nonequilibrium growth of patchy-colloid networks on
  substrates, Phys. Rev. E {\bf 87}, 032308 (2012).

\bibitem{Langevin1908} P. Langevin, Sur la th\'eorie du mouvement
  brownien, Comptes Rendue Acad. Sci. (Paris) {\bf 146}, 530 (1908).

\bibitem{Langevin} Although similar, it is not exactly the Langevin
  equation since the random forces applied to the particles are not
  due to the molecules of the medium, but, instead, to other particles
  in the medium.

\bibitem{Batchelor1967} G. K. Batchelor, {\it An Introduction to Fluid
  Dynamics} (Cambridge University Press, 1967).

\bibitem{param} The following are some values used in this work.
  Network parameters: $\sigma=1$; $m=1$; $\ell=1$; $H=15\sigma$,
  except in Fig~S2(b) where $H$ is changed; and $N=7580$, except in
  Fig.~S1 where $N$ is changed. In the particle-indenter interaction:
  $\alpha=800$ and $\epsilon=1$. The indenter radius is
  $\sigma_s=11\sigma$. The time step in the Velocity Verlet algorithm
  is 0.001.

\bibitem{Rapaport2004} D. C. Rapaport, {\it The Art of Molecular
  Dynamics Simulation} - second edition (Cambridge University Press,
  2004)
  
 \bibitem{Araujo2017} J. L. B. de Ara\'ujo, F. F. Munarin,
   G. A. Farias, F. M. Peeters, and W. P. Ferreira, Structure and
   reentrant percolation in an inverse patchy colloidal system,
   Phys. Rev. E {\bf 95}, 062606 (2017).

 \bibitem{Sneddon1965} I. N. Sneddon, The relation between load and
   penetration in the axisymmetric Boussinesq problem for a punch of
   arbitrary profile, Int. J. Eng. Sci. {\bf 3}, 47 (1965).
  
 \bibitem{Dimitriadis2002} E. K. Dimitriadis, F. Horkay, J. Maresca,
   B. Kashar, and R. S. Chadwick, Determination of elastic moduli of
   thin layers of soft material using the atomic force microscope,
   Biophys. J. {\bf 82}, 2798 (2002).

 \bibitem{Plaseied2008} A. Plaseied and A. Fatemi, Deformation
   response and constitutive modeling of vinyl ester polymer including
   strain rate and temperature effects, J. Mater. Sci. {\bf 43}, 1191
   (2008).
  
 \bibitem{Okamoto2011} R. J. Okamoto, E. H. Clayton, and P. V. Bayly,
   Viscoelastic properties of soft gels: comparison of magnetic
   resonance elastography and dynamic shear testing in the shear wave
   regime, Phys. Med. Biol. {\bf 56}, 6379 (2011).

 \bibitem{Peeters2005} E. A. G. Peeters, C. W. J. Oomens,
   C. V. C. Bouten, D. L. Bader, and F. P. T. Baaijens, Viscoelastic
   properties of single attached cells under compression,
   J. Biomech. Eng. {\bf 127}, 237 (2005).

 \bibitem{Koay2003} E. J. Koay, A. C. Shieh, and K. A. Athanasiou,
   Creep indentation of single cells, J. Biomech. Eng. {\bf 125}, 334
   (2003).


\end{thebibliography}
\end{document}